\documentclass[twocolumn]{aastex63}
\usepackage{natbib,amsmath,multirow,tabularx}

\newcommand{\dens}{g cm$^{-3}$}

\newcommand{\appropto}{\mathrel{\vcenter{
  \offinterlineskip\halign{\hfil$##$\cr
    \propto\cr\noalign{\kern2pt}\sim\cr\noalign{\kern-2pt}}}}}

\begin{document}

\title{Analytic Estimates of the Achievable Precision on the Physical Properties of Transiting Planets Using Purely Empirical Measurements}

\correspondingauthor{Romy Rodr\'iguez Mart\'inez}
\email{rodriguezmartinez.2@osu.edu}

\author[0000-0003-1445-9923]{Romy Rodr\'iguez Mart\'inez}
\affiliation{Department of Astronomy, The Ohio State University, 140 W. 18th Avenue, Columbus, OH 43210, USA}

\author[0000-0002-5951-8328]{Daniel J. Stevens}
\altaffiliation{Eberly Research Fellow}
\affiliation{Center for Exoplanets and Habitable Worlds, The Pennsylvania State University, 525 Davey Lab, University Park, PA 16802, USA}
\affiliation{Department of Astronomy \& Astrophysics, The Pennsylvania State University, 525 Davey Lab, University Park, PA 16802, USA}
\author[0000-0003-0395-9869]{B. Scott Gaudi}
\affiliation{Department of Astronomy, The Ohio State University, 140 W. 18th Avenue, Columbus, OH 43210, USA}
\author[0000-0003-3570-422X]{
Joseph G. Schulze}
\affiliation{School of Earth Sciences, The Ohio State University, 125 South Oval Mall, Columbus OH, 43210, USA}
\author[0000-0001-5753-2532]{Wendy R. Panero}
\affiliation{School of Earth Sciences, The Ohio State University, 125 South Oval Mall, Columbus OH, 43210, USA}
\author[0000-0001-7258-1834]{Jennifer A. Johnson}
\affiliation{Department of Astronomy, The Ohio State University, 140 W. 18th Avenue, Columbus, OH 43210, USA}
\affiliation{Center for Cosmology and AstroParticle Phyrics , The Ohio State University, 191 W. Woodruff Ave., Columbus, OH 43210, USA}
\author[0000-0002-4361-8885]{Ji Wang}
\affiliation{Department of Astronomy, The Ohio State University, 140 W. 18th Avenue, Columbus, OH 43210, USA}


\begin{abstract}

We present analytic estimates of the fractional uncertainties on the mass, radius, surface gravity, and density of a transiting planet, using only empirical or semi-empirical measurements. We first express these parameters in terms of transit photometry and radial velocity (RV) observables, as well as the stellar radius $R_{\star}$, if required. In agreement with previous results, we find that, assuming a circular orbit, the surface gravity of the planet ($g_p$) depends only on empirical transit and RV parameters; namely, the planet period $P$, the transit depth $\delta$, the RV semi-amplitude $K_{\star}$, the transit duration $T$, and the ingress/egress duration $\tau$. However, the planet mass and density depend on all these quantities, plus $R_{\star}$.  Thus, an inference about the planet mass, radius, and density must rely upon an external constraint such as the stellar radius. For bright stars, stellar radii can now be measured nearly empirically by using measurements of the stellar bolometric flux, the effective temperature, and the distance to the star via its parallax, with the extinction $A_V$ being the only free parameter. For any given system, there is a hierarchy of achievable precisions on the planetary parameters, such that the planetary surface gravity is more accurately measured than the density, which in turn is more accurately measured than the mass. We find that surface gravity provides a strong constraint on the core mass fraction of terrestrial planets. This is useful, given that the surface gravity may be one of the best measured properties of a terrestrial planet.

\end{abstract}

\section{Introduction}

The internal composition and structure of small, terrestrial planets is generally difficult to characterize. As is well known, mass-radius relationships alone do not constrain the internal composition of a planet beyond a measurement of its bulk density. The internal structure is crucial, as it determines the bulk physical properties of planets and provides valuable insights into their formation, history, and present composition. \citet{unterborn:2016} found that the core radius, the presence of light elements in the core, and the existence of an upper mantle have the largest effects on the final mass and radius of a terrestrial exoplanet. The final mass and radius in turn directly determine the planet's habitability. For example, the core mass fraction affects the strength of a planet's magnetic field, which shields it against harmful radiation from the host star.

At present, we have $\sim$330 small planets ($<4R_{\oplus}$) with masses and radii constrained to better than 50\%\footnote{Based on data from the NASA Exoplanet Archive, https://exoplanetarchive.ipac.caltech.edu/}. Such measurement uncertainties are generally good enough to determine the general structure of many exoplanets. However, for low-mass terrestrial planets with thin atmospheres, planetary masses and radii must be measured to precisions better than 20\% and 10\%, respectively, in order to constraint the core mass fraction and structure \citep{Dorn:2015,Schulze:2020}.

However, high-precision measurements of low-mass exoplanets between $1-4 R_{\oplus}$ are challenging. Additionally, because of the large number of individual discoveries, and because (to date) they have been mostly detected around faint \textit{Kepler}/K2 \citep{Borucki:2010,Howell:2014} targets (with typical \textit{Kepler} and K2 magnitudes of $K\sim15$ and $K\sim12$, respectively, \citealt{Vanderburg:2016}), they are difficult to follow up with high-resolution radial velocity (RV) observations and thus, obtain precise masses and other fundamental physical properties. This has already begun to change with the Transiting Exoplanet Survey Satellite mission (TESS; \citealt{Ricker:2015}), as its main science driver is to detect and measure masses and radii for at least 50 small planets ($<4 R_{\oplus}$) around bright stars. At the time of writing, 24 such planets have already been confirmed, and almost all have masses and radii measured to better than $30\%$\footnote{https://exoplanetarchive.ipac.caltech.edu/}.

The discoveries of the TESS mission will also raise very important questions in exoplanet science. The one that we address here relates to the achievable precision with which we shall be able to constrain the fundamental parameters of a transiting planet, such as its mass, density and surface gravity. Given precise photometric and spectroscopic measurements of the host of a transiting planet system, it is possible to measure the planet surface gravity with no external constraints \citep{Southworth:2007}. On the other hand, measuring the mass or radius of a transiting planet requires some external constraint \citep{Seager:2003}. Since, until very recently, it has only been possible to measure the mass or radius of the closest isolated stars directly, theoretical evolutionary tracks or empirical relations between stellar mass and radius and other properties of the star have often been used (e.g., \citealt{Torres:2010}).  However, these constraints typically assume that the star is representative of the population of systems that were used to calibrate these relations.  In the case of theoretical evolutionary tracks, there may be systematic errors due to uncertainties in the physics of stellar structure, atmospheres, and evolution, or second-order properties of the star, such as its detailed abundance distribution, which can manifest as irreducible systematic uncertainties on the stellar parameters. For example, most evolutionary tracks assume a fixed solar abundance pattern scaled to the iron abundance [Fe/H] of the star, and thus, the same [$\alpha$/Fe] as the Sun. If the host star has a significantly different [$\alpha$/Fe] than the Sun, that will lead to incorrect inferences about the properties of the planet. By using evolutionary tracks that assume a solar [$\alpha$/Fe], one might infer an incorrect density and mass of the planet, and therefore an incorrect core/mantle fraction. 

Thus, a direct, empirical or nearly empirical measurement of the radius or mass of the star that does not rely on assumptions that may not be valid is needed (see \citealt{Stassun2017} for a lengthier discussion on the merits and benefits of using empirical or semi-empirical measurements to infer exoplanet parameters). As has been demonstrated in numerous papers (see, e.g., \citealt{Stevens:2017}), with \textit{Gaia} \citep{Gaia:2018} parallaxes, coupled with the availability of absolute broadband photometry from the near-UV to the near-IR, it is now possible to measure the radii of bright ($V\la 12$ mag) stars. This allows for direct, nearly empirical measurements of the masses and radii of transiting planets and their host stars (and, indeed, any eclipsing single-lined spectroscopic binary) in a nearly empirical way. See \citet{Stevens:2018} for an initial exploration of the precisions with which these measurements can be made. 

In this paper, we build upon the work of \citet{Stevens:2018} by also assessing the precision with which the surface gravity $g_p$ of transiting planets can be measured. Given that a measurement of $g_p$ only requires measurements of direct observables from the transit photometry and radial velocities without the need for external constraints, the precision on $g_p$ in principle improves with ever more data, assuming no systematic floor. Thus we seek to address two questions.  First, with what fractional precision can $g_p$ be measured, and how does this compare to the fractional precision with which the density or mass can be measured?  Second, how useful is $g_p$ as a diagnostic of a terrestrial planet's interior structure and, potentially, habitability?

Answering these questions is quite important because the surface gravity of a planet may be a more fundamental parameter than the radius and mass, at least in addressing certain questions, such as its habitability \citep{oneill:2007, valencia:2009,vanHeck:2011}.  For example, the surface gravity, along with the equilibrium temperature and mean molecular weight, determines the scale height of any extant atmosphere. If a planet's surface gravity provides more of a lever arm in determining certain aspects of the planet's interior or atmosphere, and if we can achieve a better precision on the planet surface gravity measurement than the radius, then we can use that to better constrain the composition of the planet and, ultimately, its habitability. Thus, given the importance of the planetary surface gravity, mass and radius in constraining the habitability of a planet, it is critical to understand how well we can measure these properties.

Here we focus on the precision with which the surface gravity, density, mass, and radius of a transiting planet can be measured essentially empirically. We will employ methodologies that are similar to those used in \citet{Stevens:2018}, and thus this work can be considered a companion paper to that one.  

\section{Analysis}
\label{sec:analysis}

We begin by deriving expressions for the surface gravity $g_p$, mass $M_p$, density $\rho_p$, and radius $R_p$, of a transiting planet in terms of observables from photometric and radial velocity observations, as well as a constraint on the stellar radius $R_{\star}$ from a \textit{Gaia} parallax combined with a bolometric flux from a spectral energy distribution (SED) \citep{Stassun2017,Stevens:2017,Stevens:2018}.  

\subsection{Planet Surface Gravity\label{sec:gp}}

The planet surface gravity is defined as
\begin{equation}
    g_{p} = \frac{GM_{p}}{R_{p}^{2}}.
\end{equation}

The radial velocity semi-amplitude $K_{\star}$ can be expressed as 
\begin{equation}
\begin{split}
    K_{\star} =\left(\frac{2\pi G}{P}\right)^{1/3} \frac{M_p\sin{i}}{(M_{\star}+M_p)^{2/3}} \frac{1}{\sqrt{1-e^2}} \\
      \simeq28.4~{\rm m~s}^{-1}\left(\frac{P}{\rm yr}\right)^{-1/3} \frac{M_{p}\sin{i}}{M_{\rm J}}\bigg(\frac{M_{\star}}{M_{\odot}}\bigg)^{-2/3}
      (1-e^2)^{-1/2},
    \end{split}
    \label{eqn:K}
\end{equation}

where $M_\star$ is the stellar mass, $P$ and $e$ are the planetary orbital period and eccentricity, $M_{\rm J}$ is Jupiter's mass, and $i$ is the inclination angle of the orbit.  In the second equality, we have assumed that $M_p \ll M_{\star}$.

Using Newton's version of Kepler's third law and Equation~\ref{eqn:K}, the surface gravity can then be expressed as
\begin{equation}
    g_{p} =\frac{2\pi}{P}
    \frac{\sqrt{1-e^2}}{(R_p/a)^2}
    \frac{K_{\star}}{\sin{i}}.
\end{equation}

For the majority of the following analysis, will assume circular orbits ($e=0$) for simplicity and drop the eccentricity dependence.  This analysis could be repeated for eccentric orbits, but the algebra is tedious and does not lead to qualitatively new insights. The assumption of circular orbits thus provides a qualitative expectation of the uncertainties on the planetary parameters.  Furthermore, in many cases it is justified because we expect many of the systems to which this analysis is applicable will have very small eccentricities. We also further assume that $\sin{i}=1$, which is approximately true for transiting exoplanets.  Under these assumptions, we have that
\begin{equation}
    g_{p} =\frac{2\pi K_{\star}}{P}
    \left(\frac{a}{R_p}\right)^2.
    \label{eqn:gp}
\end{equation}

The semimajor axis scaled to the planet radius can be converted to the semimajor axis scaled to the stellar radius by using the depth of the transit $\delta\equiv (R_p/R_{\star})^2$, which is a direct observable:
\begin{equation}
    \frac{a}{R_{p}} = \frac{a}{R_{\star}}\left(\frac{R_{\star}}{R_{p}}\right) = \frac{a}{R_{\star}}\delta^{-1/2}.
    \label{eqn:arparstarphys}
\end{equation}
We can then rewrite the scaled semimajor axis $a/R_{\star}$ in terms of the stellar density $\rho_{\star}$ (see e.g., \citealt{Sandford:2017} for a precise derivation):

\begin{equation}
    \frac{a}{R_{\star}} = 
    \left(\frac{GP^2}{3\pi}\right)^{1/3}
    \left(\rho_{\star}+k^3\rho_p\right)^{1/3},
    \label{eqn:arstar}
\end{equation}
where $k\equiv R_p/R_{\star}$.  Since typically $k\ll 1$, 
\begin{equation}
    \frac{a}{R_{\star}} = \left(\frac{GP^2\rho_{\star}}{3\pi}\right)^{1/3}.
    \label{eqn:arstar8}
\end{equation}
Using Equation \ref{eqn:arparstarphys}, we find
\begin{equation}
    \frac{a}{R_p} = \left(\frac{GP^2\rho_{\star}}{3\pi}\right)^{1/3}\delta^{-1/2}.
\end{equation}

Noting that, from Equation \ref{eqn:gp}, we can write
\begin{equation}
    g_{p} = \frac{2\pi K_{\star}}{P}
    \left(\frac{a}{R_p}\right)^2=\frac{2\pi K_{\star}}{P}
    \left(\frac{a}{R_{\star}}\right)^2\delta^{-1}.
\label{eqn:gp2}
\end{equation}
Then $a/R_{\star}$ can be written in terms of observables as
\begin{equation}
\frac{a}{R_{\star}}= \frac{P}{\pi}\frac{\delta^{1/4}}{\sqrt{T\tau}}.
\label{eqn:arstarobs}
\end{equation}
where the observables are the orbital period $P$, transit time $T$ (full-width at half-maximum), the ingress/egress duration $\tau$, and the transit depth $\delta$. 

Inserting this into Equation \ref{eqn:gp2}, we find that the planet surface gravity is given in terms of pure observables as (\citealt{Southworth:2007})
\begin{equation}
    g_{p} = \frac{2K_{\star}P}{\pi T\tau} \delta^{-1/2}.
\label{eqn:gpobservables}
\end{equation}

Using linear propagation of uncertainties, and assuming no covariances between the observable parameters\footnote{See \citet{Carter:2008} for an exploration of the covariances between photometric and RV observable parameters.}, and the aforementioned assumptions ($M_P\ll M_{\star}$, $\sin{i}\sim 1$, $e=0$, {\rm and} $k\ll 1$), we can approximate the fractional uncertainty on the surface gravity as

\begin{equation}
\begin{split}
\bigg(\frac{\sigma_{g_p}}{g_p}\bigg)^{2}     \approx \bigg(\frac{\sigma_{K_{\star}}}{K_{\star}}\bigg)^{2} +  
    \bigg(\frac{\sigma_{P}}{P}\bigg)^{2} +
    \bigg(\frac{\sigma_{T}}{T}\bigg)^{2} +  \\
    \bigg(\frac{\sigma_{\tau}}{\tau}\bigg)^{2} +
    \frac{1}{4}\bigg(\frac{\sigma_{\delta}}{\delta}\bigg)^{2}.
    \end{split}
    \label{eqn:gpuncertianty}
\end{equation}

\subsection{Planet Mass\label{sec:mp}}

We now turn to the uncertainty on the planet mass.  We can approach this estimate two ways.  First, we can start from Equation \ref{eqn:K}, again making the same simplifying assumptions, and solve for $M_p$ in terms of observables.  We note that this method requires the intermediate step of deriving an expression for the host star mass in terms of direct observables, using the fact that
\begin{equation}
M_{\star}=\frac{4\pi}{3}\rho_{\star}R_{\star}^3
\label{eqn:mstarphys}
\end{equation}
and using Equations \ref{eqn:arparstarphys} and \ref{eqn:arstarobs} to write $\rho_{\star}$ in terms of observables. We find
\begin{equation}
    M_{\star}=\frac{4P}{\pi G}\delta^{3/4}(T\tau)^{-3/2}R_{\star}^3.
\end{equation}
Using this, we can then derive the planet mass in terms of observables as 
\begin{equation}
M_p = \frac{2}{\pi G}\frac{K_{\star}P}{T\tau}R_{\star}^2\delta^{1/2}.
\label{eqn:mpobs}
\end{equation}

A more straightforward approach is to use the fact that we have already derived the planet surface gravity in terms of observables.  Starting from the definition of surface gravity, we can write
\begin{equation}
M_p=\frac{1}{G}g_p R_p^2=\frac{1}{G}g_p R_{\star}^2\delta.
\end{equation}
Using Equation \ref{eqn:gpobservables}, we arrive at the same expression as Equation \ref{eqn:mpobs}.

Using Equation \ref{eqn:mpobs}, we derive the fractional uncertainty on the planet mass in terms of the fractional uncertainty in the observables, again assuming no covariances and the simplifying assumptions stated before ($M_P\ll M_{\star}$, $\sin{i}\sim 1$, $e=0$, {\rm and} $k\ll 1$).  We find
\begin{equation}
\begin{split}
   \bigg(\frac{\sigma_{M_p}}{M_p}\bigg)^{2} \approx \bigg(\frac{\sigma_{K_{\star}}}{K_{\star}}\bigg)^{2} +  \bigg(\frac{\sigma_{P}}{P}\bigg)^{2} +
    \bigg(\frac{\sigma_{T}}{T}\bigg)^{2} +  \\
    \bigg(\frac{\sigma_{\tau}}{\tau}\bigg)^{2} +  
    \frac{1}{4}\bigg(\frac{\sigma_{\delta}}{\delta}\bigg)^{2} +
    4\bigg(\frac{\sigma_{R_{\star}}}{R_{\star}}\bigg)^{2}.
\end{split}
\label{eqn:mpuncertainty}
\end{equation}

\subsection{Planet Density\label{sec:density}}

We derive the planet density $\rho_{p}$ in terms of observables.  The planet density is given by
\begin{equation}
    \rho_p = \frac{3M_p}{4\pi R_p^3}=\frac{3M_p}{4\pi R_{\star}^3}\delta^{-3/2}.
\end{equation}
We have already derived the mass of the planet in terms of observables in Equation \ref{eqn:mpobs}.  Using this expression, we find
\begin{equation}
    \rho_p = \frac{3}{2\pi^2G}\frac{K_{\star}P}{T\tau\delta R_{\star}}.
    \label{eqn:rhopobs}
\end{equation}
From this equation, we derive the fractional uncertainty on the planet density in terms of the fractional uncertainty in the observables, again assuming no covariances and the simplifying assumptions stated before.  We find
\begin{equation}
\begin{split}
   \bigg(\frac{\sigma_{\rho_p}}{\rho_p}\bigg)^{2} \approx \bigg(\frac{\sigma_{K_{\star}}}{K_{\star}}\bigg)^{2} +  \bigg(\frac{\sigma_{P}}{P}\bigg)^{2} +
    \bigg(\frac{\sigma_{T}}{T}\bigg)^{2} +  \\
     \bigg(\frac{\sigma_{\tau}}{\tau}\bigg)^{2} +  
    \bigg(\frac{\sigma_{R_{\star}}}{R_{\star}}\bigg)^{2} +
    \bigg(\frac{\sigma_{\delta}}{\delta}\bigg)^{2}.
\end{split}
\label{eqn:rhopuncertainty}
\end{equation}

\subsection{Planet Radius \label{sec:Rp}}

Finally, the planet radius uncertainty can be trivially derived from the definition of the transit depth $\delta$, assuming no limb darkening: 

\begin{equation}
    \delta = \bigg(\frac{R_{p}}{R_{\star}}\bigg)^{2}.
    \label{eqn:deltadef}
\end{equation}

Then, 
\begin{equation}
    R_{p} = \sqrt{\delta}R_{\star},
    \label{eqn:rpobs}
\end{equation}

and the fractional uncertainty on the planet radius is simply

\begin{equation}
   \bigg(\frac{\sigma_{R_{p}}}{R_{p}}\bigg)^{2} \approx
     \bigg(\frac{\sigma_{R_{\star}}}{R_{\star}}\bigg)^{2} +
    \frac{1}{4}\bigg(\frac{\sigma_{\delta}}{\delta}\bigg)^{2}.
    \label{eqn:runcertainty}
\end{equation}

We note that, by assuming that $\delta$ is a direct observable, we are fundamentally assuming no limb darkening of the star.  Of course, in reality the presence of limb darkening means that the observed fractional depth of the transit is not equal to $\delta$, and thus the uncertainty in $\delta$ is larger than one would naively estimate assuming no limb darkening.  However, assuming that the limb darkening is small (as it is for observations in the near-IR), or that it can be estimated a priori based on the properties of the star, or that the photometry is sufficiently precise that both the limb darkening and $\delta$ can be simultaneously constrained, the naive estimate of the uncertainty on $\delta$ assuming no limb darkening will not be significantly larger than that in the presence of limb darkening. 

\section{Comparing the Estimated Uncertainties on the Planet Mass, Density, and Surface Gravity}

Comparing the expressions for the fractional uncertainty on $g_p$, $M_p$, and $\rho_p$  (Equations ~\ref{eqn:gpuncertianty},~\ref{eqn:mpuncertainty}, and~\ref{eqn:rhopuncertainty}, respectively), we can make some broad observations on the precision with which it is possible to measure these three planetary parameters.  

First, comparing the uncertainties on $g_p$ and $M_p$, we note that the only difference is that $\sigma_{M_{p}}/M_{p}$ requires the additional term $4(\sigma_{R_{\star}}/R_{\star})^2$.  \citet{Stevens:2018} estimates that it should be possible to infer the stellar radii of bright hosts ($G\la 12$ mag) to an accuracy of order $1\%$ using the final \textit{Gaia} data release parallaxes, currently-available absolute broadband photometry, and spectrophotometry from \textit{Gaia} and the Spectro-Photometer for the History of the Universe, Epoch of Reionization, and Ices Explorer (SPHEREx; \citealt{SPHEREx:2018}). The exact level of accuracy will depend on the stellar spectral type and the final parallax precision. It is likely that $R_{\star}$ may dominate the error budget relative to the other terms, with the possible exception of the uncertainty in $\tau$.  We note that TESS is able to measure $\tau$ more precisely than either {\it Kepler} or K2 were able to for systems with similar physical parameters and noise properties, primarily because the TESS bandpass is redder than that of {\it Kepler}, and thus the stellar limb darkening is smaller and less degenerate with $\tau$.  Overall, we generically expect the planetary surface gravity to be measured to smaller fractional precision than the planet mass. 

We now turn to the uncertainty on planetary density.  When comparing the expressions for the uncertainty in $M_p$ to $\rho_p$, we note that the uncertainty due to the depth enters as $(1/4)(\sigma_\delta/\delta)$ for $M_p$, whereas it enters as simply $\sigma_\delta/\delta$ for $\rho_p$.  For large planets, the depth should be measurable to a precision of $\sim$1\% or better, particularly in the TESS bandpass, similar to the best expected precision on $R_{\star}$.  Thus, we expect $\sigma_{\delta}$ to be comparable to $\sigma_{R_{\star}}$, and thus both should contribute at the $\sim$1\% level to $\sigma_{\rho_p}$.  On the other hand, we expect $\sigma_{R_{\star}}$ to dominate over the transit depth for $M_p$. Thus, for any given system, we generally expect the following hierarchy: $\sigma_{M_p}/M_p>\sigma_{\rho_p}/\rho_p>\sigma_{g_p}/\sigma_{g_p} > \sigma_{R_p}/R_p$.

Similarly, there is a hierarchy in the precision with which the observed parameters $T$, $P$, $K_{\star}$, $\delta$, $\tau$, and $R_{\star}$ are measured. For the relatively small sample of planets confirmed from TESS so far, we find that in general, the most precise observable parameter is the orbital period, followed by the stellar radius, the transit depth, the RV semi-amplitude, and the transit duration, such that:    $\sigma_{T}/T>\sigma_{K}/K>\sigma_{\delta}/\delta>\sigma_{R_{\star}}/R_{\star} >\sigma_{P}/P$. The ingress/egress time $\tau$ is not always reported in discovery papers, so we do not include it in this comparison. However, we generally expect that it will be measured to a precision that is worse than $T$ \citep{Carter:2008, Yee:2008}. 

This hierarchy is in agreement with the findings of \citet{Carter:2008} and \citet{Yee:2008}, who derived the following approximate relations for the uncertainties in the parameters of a photometric transit (assuming no limb darkening):
\begin{eqnarray}
\frac{\sigma_{\delta}}{\delta} &\simeq& Q^{-1}\\
\frac{\sigma_T}{T} &\simeq& Q^{-1}\sqrt{\frac{2\tau}{T}}\\
\frac{\sigma_\tau}{\tau} &\simeq& Q^{-1}\sqrt{\frac{6T}{\tau}},
\label{eqn:transuncertainties}
\end{eqnarray}

where $Q$ is the signal-to-noise ratio of the combined transits, defined as
\begin{equation}
    Q\equiv(N_{\rm tr}\Gamma_{\rm phot}T)^{1/2}\delta,
\label{eqn:Q}
\end{equation}
where $N_{\rm tr}$ is the effective number of transits that were observed, and $\Gamma_{\rm phot}$ is the photon collection rate\footnote{Alternatively, assuming all measurements have a fractional photometric uncertainty $\sigma_{\rm phot}$, and there are $N$ measurements in transit, the total signal-to-noise ratio can be defined as $Q\equiv \sqrt{N}(\delta/\sigma_{\rm phot})$.}. We note that Equation \ref{eqn:Q} implicitly assumes uncorrelated photometric uncertainties. Since, in general, $\sigma_{\tau} > \sigma_{T}$, we have that $\sigma_{\delta}/\delta<\sigma_{T}/T<\sigma_{\tau}/\tau$.

In the above equations, we have ignored the uncertainty in the transit midpoint $t_c$ as it does not enter into the expressions for the uncertainties in $R_p$, $M_p$, $\rho_p$, or $g_p$. We also assumed that the uncertainty in the baseline (out of transit) flux is negligible, which is generally a good assumption, particularly for space-based missions such as {\it Kepler}, K2, and TESS, where the majority of the measurements are taken outside of transit.  

We note that, particularly for small planets, when the limb darkening is significant, and when only a handful of transits have been observed, $\tau$ may be poorly measured (i.e., precisions of $\ga 5\%$), and therefore its uncertainty will dominate the error budget.  In such cases, it may be more prudent to use additional external constraints, such as stellar isochrones, to improve the overall parameters of the system (at the cost of losing the nearly purely empirical nature of the inferences as assumed in the derivations above).  See \citet{Stevens:2018} for additional discussion.

\section{Validation of Our Analytic Estimates}
\label{validation}

\begin{figure}
\begin{center}
\includegraphics[width=\columnwidth]{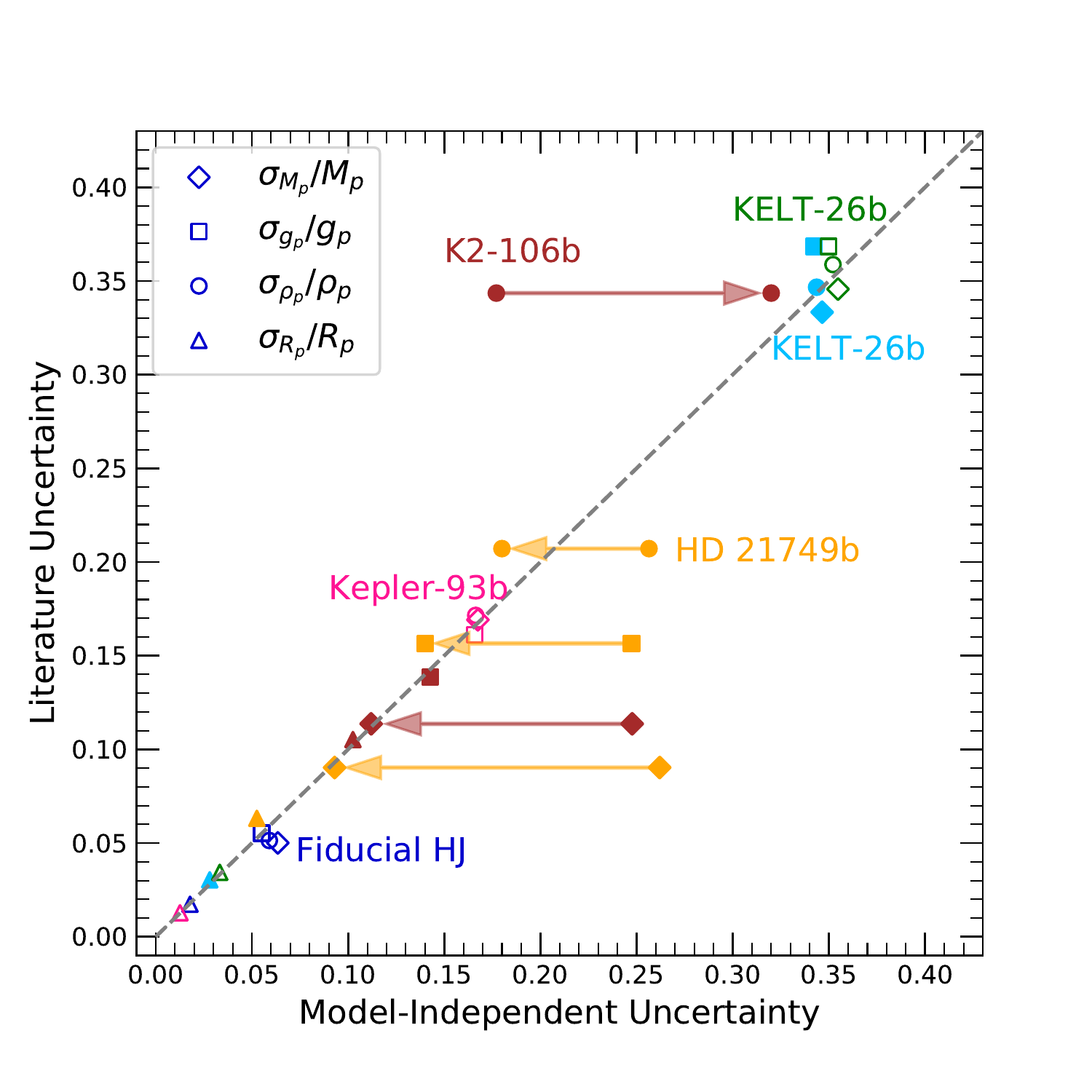}
\end{center}
\vspace{-3mm}
\caption{Reported fractional uncertainties in $M_{p}$ (diamonds), $g_{p}$ (squares),  $\rho_{p}$ (circles), and $R_{p}$ (triangles) versus our model-independent analytic estimates for a variety of transiting planets.  These include a fiducial hot Jupiter (dark blue), Kepler-93b (pink), KELT-26b (sky blue), KELT-26b without external constraints (green, open symbols),  K2-106b (brown), and HD 21749b (gold). For K2-106b and HD 21749b, arrows point from the fractional uncertainties reported in the discovery papers to our `forensic' estimates of the uncertainties that could be achieved had the authors adopted only empirical constraints. The open symbols are systems for which no external constraints were used. A dashed, gray one-to-one line is plotted for reference.}
\label{fig:fig1}
\end{figure}

We test the analytic expressions derived in Section~\ref{sec:analysis} using four confirmed exoplanets and one fiducial hot Jupiter simulated by \citet{Stevens:2018}. The confirmed exoplanets are KELT-26b \citep{Rodriguez:2020}, HD 21749b \citep{Dragomir:2019}, K2-106b \citep{Adams:2017,Guenther:2017}, and Kepler-93b \citep{Ballard:2014,Dressing:2015}. These systems have masses and radii between $\sim$4--448 $M_{\oplus}$ and $\sim$1.4--21 $R_{\oplus}$. 

We estimated the expected analytic uncertainties on the planet parameters by inserting the values of $T$, $P$, $K_{\star}$, $\delta$, $\tau$, $R_{\star}$, and their respective uncertainties from the discovery papers into Equations~\ref{eqn:gpuncertianty},~\ref{eqn:mpuncertainty}, \ref{eqn:rhopuncertainty}, and~\ref{eqn:runcertainty}. Then, we compared the analytic uncertainties to those reported in the discovery papers, which were derived from MCMC analyses and not using the analytic approximations presented here. For parameters with asymmetric uncertainties, we took the average of the upper and lower bounds and adopted that as the uncertainty.

We note that the discovery papers of all of the examples we present here (except for KELT-26b) do not provide the transit duration $T$ (or the full width at half maximum of the transit), but rather $T_{14}$, which is defined as the difference between the fourth and first contact (see, e.g., \citealp{Carter:2008}). Since we are generally interested in $T$, we calculate it from the given observables using
\begin{equation}
    T = T_{14} - \tau 
    \label{eqn:T}
\end{equation}
and we estimate its uncertainty with the relationship from \citet{Carter:2008} and \citet{Yee:2008}:
\begin{equation}
    \frac{\sigma_{T}}{T} =\left(\sqrt{\frac{1}{3}}\frac{\tau}{T}\right)\frac{\sigma_{\tau}}{\tau}.
    \label{eqn:sigmaT}
\end{equation}
We use Equations~\ref{eqn:T} and~\ref{eqn:sigmaT} to calculate the transit duration values and uncertainties for all the systems for which only $T_{14}$ is given.

There are other ways of estimating the uncertainty in $T$, such as by propagating the uncertainty on $T$ from Equation~\ref{eqn:T}, or by assuming that the uncertainty in $T$ is approximately equal to that of $T_{14}$. However, these approaches overestimate the uncertainty on $T$ as compared to Equation~\ref{eqn:sigmaT} because they do not account for the covariance between the measurements of $T_{14}$ and $\tau$.  Therefore, we adopt the uncertainty in $T$ from Equation~\ref{eqn:sigmaT} for all the exoplanets referenced here.

Finally, for systems where the transit depth $\delta$ is not provided, but rather the planet-star radius ratio $R_{p}/R_{\star}$, we use linear propagation of error to estimate $\sigma_{\delta}$, finding
\begin{equation}
    \bigg(\frac{\sigma_{\delta}}{\delta}\bigg)^{2} = 4\bigg(\frac{\sigma_{R_{p}/R_{\star}}}{R_{p}/R_{\star}}\bigg)^{2}.
    \label{eqn:uncrpvsdelta}
\end{equation}

And we adopt the fractional uncertainties on $R_{\star}$ as reported in the papers. In some cases these were derived using external constraints, such as stellar models, and thus may be underestimates or overestimates of the empirical uncertainty in $R_{\star}$ derived from the stellar SED and parallax. 

The fractional uncertainties calculated using our analytic approximations for the five planets in our sample are listed in Table~\ref{tab:analytic_numerical} and shown in Figure~\ref{fig:fig1}. As is clear from Figure~\ref{fig:fig1}, our estimates are broadly in agreement with the fractional uncertainties quoted in the discovery papers. However, we note that the fractional uncertainties we predict for certain quantities are systematically larger or smaller than those reported in the papers. After a careful `forensic' analysis, we have tracked down the reason for these discrepancies.  In the two most discrepant cases, it is because the authors used external constraints on the properties of the host star (such as $T_{\rm eff}$, [Fe/H], and $\log{g_\star}$) combined with stellar evolutionary tracks, to place priors on the stellar parameters $R_\star$ and $M_\star$. In one case (HD 21749), the resulting constraint on $\rho_{\star}$ is {\it tighter} than results from empirical constraint on the stellar density $\rho_{\star}$ from the light curve.  In the other case, the adopted constraint on $\rho_{\star}$ is {\it weaker} than results from empirical constraint on the stellar density $\rho_{\star}$ from the light curve, but nevertheless the external (weaker) constraints on $M_{\star}$ and $R_{\star}$ were adopted, rather than the (tighter) empirical constraints. In the remaining cases, either external constraints were not assumed, and as a result their parameter uncertainties agree well with our analytic estimates, or the external constraints were negligible compared to the empirical constraints and thus the empirical constraints dominated, again leading to agreement with our analytic estimates.  We ultimately conclude that our analytic estimates are reliable; however, we describe in detail our forensic analysis of the systems for pedagogical purposes.  In the following subsections, we discuss each system in further detail.

Before doing so, however, we stress that the advantage of empirical, model-independent approximations like the ones presented here is that they do not assume that the physical properties of any particular system is representative of the systems used to calibrate the empirical models, or that the properties of the systems necessarily agree with the theoretical predictions.  For example, theoretical models that make assumptions about the elemental abundances of the host star may not apply to the particular system under consideration.  Therefore, although our empirical approach may lead to weaker constraints on the parameters of the planets, we believe it leads to more robust constraints on these parameters.

\setlength{\tabcolsep}{4pt}
\begin{deluxetable*}{lcccc|ccccc}
\tablecolumns{10}
\tablecaption{Analytic and reported fractional
uncertainties\label{tab:analytic_numerical}}
\tablehead{\multicolumn{1}{c}{\textbf{Planet}} & \multicolumn{4}{c|}{\textbf{\textbf{Analytic}}}  & \multicolumn{4}{c}{\textbf{Literature}} & \multicolumn{1}{c}{\textbf{Reference}}\\
\cline{2-5}
\cline{6-9}
\colhead{} &  \colhead{$M_p$} & \colhead{$g_p$} & \colhead{$\rho_p$} &\multicolumn{1}{c|}{$R_p$} & \colhead{$M_p$} & \colhead{$g_p$} & \colhead{$\rho_p$} & \colhead{$R_p$}  & \colhead{}}

\startdata
Fiducial HJ & 0.06 & 0.04 & 0.06 &0.01& 0.05 & 0.05 & 0.05 &0.01 & \citet{Stevens:2018}  \\
KELT-26b & 0.34 & 0.34 & 0.34 &0.03 & 0.33 & 0.37 & 0.35 &0.03& \citet{Rodriguez:2020} \\
KELT-26b$^{\star}$ & 0.35 & 0.35 & 0.35 &0.03 & 0.33 & 0.37 & 0.35 &0.03& \\
Kepler-93b & 0.17	& 0.16 & 0.17 & 0.01& 0.17 & 0.16 &0.17 &0.01 & \citet{Ballard:2014} \\
HD 21749b$^{\dagger}$ & 0.26 & 0.25 & 0.26 &0.05 & 0.09 & 0.16 & 0.21 & 0.06& \citet{Dragomir:2019}\\
&(0.09) & (0.14) &  (0.18) & & & & & \\
K2-106b$^{\dagger}$ & 0.25 & 0.14 & 0.18 & 0.10& 0.11 & 0.13 & 0.34 & 0.10& \citet{Guenther:2017}\\
&(0.11) & &  (0.32) & & & & & \\
\enddata
\tablenotetext{}{\textbf{Notes.} The first four columns are the analytic uncertainties in $M_{p}$ (Eqn~\ref{eqn:mpuncertainty}), $g_{p}$ (Eqn~\ref{eqn:gpuncertianty}), $\rho_{p}$ (Eqn~\ref{eqn:rhopuncertainty}), and $R_{p}$ (Eqn~\ref{eqn:runcertainty}), while the next four are the uncertainties in those parameters reported in the literature. Planets with a $^{\dagger}$ were analyzed using external constraints from stellar evolutionary models. KELT-26b$^{\star}$ is KELT-26 analyzed without external constraints. The quantities in parentheses below HD 21749b and K2-106b are the values we recover if we assume external constraints, as explained in Sections~\ref{sec:hd21749b} and~\ref{sec:k2106}.} 
\end{deluxetable*}

\subsection{A Fiducial Hot Jupiter}

\citet{Stevens:2018} simulated the photometric time series and RV measurements for a typical hot Jupiter ($M_{p}= M_{\rm J}$ and $R_{p}= R_{\rm J}$) on a 3 day orbit transiting a G-type star using VARTOOLS \citep{Hartman:2016}. They injected a Mandel-Agol transit model \citep{Mandel:2002} into an (out-of-transit flux) normalized light curve, and simulated measurement offsets by drawing from a Gaussian distribution with 1 millimagnitude dispersion.  They furthermore assumed a cadence of 100 seconds. They note that these noise properties are typical of a single ground-based observation of a hot Jupiter from a small, $\sim$1 m telescope. For the RV data, they simulated 20 evenly-spaced measurements, each with 10 m $\rm s^{-1}$ precision (which they assumed was equal to the scatter, or `jitter'). They then performed a joint photometric and RV fit to the simulated data using EXOFASTv2 \citep{Eastman:2017,eastman:2019} to model and estimate the star and planet's properties. They simulated three different cases: a circular ($e=0$) orbit and equatorial transit, or an impact parameter of $b=0$, an eccentric orbit with $e=0.5$ and $b=0$, and a circular orbit and $b=0.75$. We consider the parameters and uncertainties for the case of a circular orbit and equatorial transit, for which our equations are most applicable, and use the best-fit values and uncertainties from Table 1 in \citet{Stevens:2018}. 

The fractional uncertainties in the planet mass, surface gravity, and planetary bulk density quoted in \citet{Stevens:2018} are all roughly 5\%, whereas the fractional uncertainty in the planet radius is 1.7\%. These uncertainties are in very good agreement with our analytic estimates, as Figure~\ref{fig:fig1} and Table~\ref{tab:analytic_numerical} show. 

\subsection{A Real Hot Jupiter}

KELT-26b is an inflated ultra hot Jupiter on a 3.34 day, polar orbit around a early Am star characterized by \citet{Rodriguez:2020}. It has a mass and radius of $M_{p}=1.41^{+0.43}_{-0.51} M_{\rm J}$ and $R_{p}=1.940^{+0.060}_{-0.058} R_{\rm J}$, respectively. The photometry (which included TESS data) and radial velocity data were jointly fit using EXOFASTv2, and included an essentially empirical constraint on the radius of the star from the spectral energy distribution and the \textit{Gaia} Data Release 2 (DR2) parallax, as well as theoretical constraints from the MESA Isochrones and Stellar Tracks (MIST) stellar evolution models \citep{Dotter:2016,Choi:2016,Paxton:2011,Paxton:2013,Paxton:2015}. Therefore, unlike the fiducial hot Jupiter discussed above, this system was modeled using both external empirical constraints and external theoretical constraints. The uncertainties in the planet parameters reported by \citet{Rodriguez:2020} are $\sim$34\% for the mass, $\sim$33\% for the surface gravity, $\sim$33\% for the bulk density, and 3.8\% for the planet's radius.  These are very close to our estimates of the fractional uncertainties of these parameters, implying that the constraints from the MIST evolutionary tracks have little effect on the inferred parameters of the system. 

To test this hypothesis, we reanalyzed this system with EXOFASTv2 without using the external theoretical constraints from the MIST isochrones, that is, only using the spectral energy distribution of the star, its parallax from \textit{Gaia} DR2, and the light curves and radial velocities. The uncertainties from this analysis are 35\% for the planetary mass, surface gravity, and the density, and 3.3\% for the radius. These are consistent with the uncertainties derived from the analysis using the MIST evolutionary tracks as constraints.  The fractional uncertainties from the original paper and the analysis without constraints are shown in sky blue (with constraints) and green (without) in Figure~\ref{fig:fig1}.  We conclude that the inferred parameters of the system derived using purely empirical constraints are as precise (and likely more accurate) than those inferred using theoretical evolutionary tracks.  Therefore, at least for systems similar to KELT-26, we see no need to invoke theoretical priors.  

\subsection{Kepler-93b}

Kepler-93b is a terrestrial exoplanet on a 4.7 day period discovered by \citet{Ballard:2014}. It has a mass and radius of $M_{p}= 4.02 \pm0.68~M_{\oplus}$ and $R_{p}= 1.483\pm0.019~R_{\oplus}$. With a radius uncertainty of only 1.2\%, it is one of the most precisely characterized exoplanets to date. \citet{Ballard:2014}
used asteroseismology to precisely constrain the stellar density, and then used it as a prior in their MCMC analysis, leading to the remarkably precise planet radius. Their analysis did not use external constraints from stellar evolutionary models, however.  \citet{Dressing:2015} revisited Kepler-93 and collected HARPS-N \citep{mayor:2003} spectra, which they combined with archival Keck/HIRES spectra to improve upon the planet's mass estimate. They thus reduced the uncertainty in the mass of Kepler-93b from $\sim$40\% \citep{Ballard:2014} to $\sim$17\%. We used the photometric parameters ($T$, $\tau$, and $\delta$) from \citet{Ballard:2014} and the semi-amplitude $K_{\star}$ from \citet{Dressing:2015} to test our analytic estimates. We compared our results to the reported uncertainties in $M_{p}$, $g_{p}$ and $\rho_{p}$ from \citet{Dressing:2015}, since they provide slightly more precise properties. The uncertainties in the properties of Kepler-93b are all $\sim$17\%, and 1.2\% for the radius, which are in excellent agreement with our analytic estimates, as shown in Figure~\ref{fig:fig1} and Table~\ref{tab:analytic_numerical}.  Interestingly, this implies that the asteroseismological constraint on $\rho_{\star}$ does not significantly improve the overall constraints on the system.  

\subsection{HD 21749b}
\label{sec:hd21749b}

HD 21749b is a warm sub-Neptune on a 36 day orbit transiting a K4.5 dwarf discovered by \citet{Dragomir:2019}. The planet has a radius of $2.61^{+0.17}_{-0.16} R_{\oplus}$ determined from TESS data, and a mass of $22.7^{+2.2}_{-1.9} M_{\oplus}$ constrained from high-precision, radial velocity data from the HARPS spectrograph at the La Silla Observatory in Chile. \citet{Dragomir:2019}
performed an SED fit combined with a parallax from \textit{Gaia} DR2
to constrain the host star's radius to $R_{\star} = 0.695\pm 0.030 R_{\odot}$. They then used the \citet{Torres:2010} relations to derive a stellar mass of $M_{\star} =0.73\pm0.07 M_{\odot}$, although they do not specify what values of $T_{\rm eff}$, [Fe/H], and $\log{g_\star}$ they adopt as input into those equations, or from where they derive these values.  We assume they were determined from high-resolution stellar spectra.  Finally, they performed a joint fit of their data and constrained the planetary parameters with the EXOFASTv2 modeling suite, using their inferred values of $M_{\star}$ and $R_{\star}$ as priors. 

When comparing our analytic approximations of the fractional uncertainties in $M_{p}$, $g_{p}$, and $\rho_{p}$ to the uncertainties in the paper, we find that our estimates are systematically larger than those of \citet{Dragomir:2019} by 34\% ($M_{p}$), 60\% ($g_{p}$), and 80\% ($\rho_{p}$). 

Understanding the nature of such discrepancies requires a closer examination of the methods employed by \citet{Dragomir:2019} as compared to ours. The fundamental difference is that their uncertainties in the planetary properties are dominated by their more precise {\it a priori} uncertainties on $M_{\star}$ and $R_{\star}$ (and thus $\rho_{\star}$), rather than the empirically constrained value of $\rho_{\star}$ from the light curve and radial velocity measurements. On the other hand, we estimate the uncertainty on $\rho_{\star}$ directly from observables (e.g., the light curve and the RV data).  

Because their prior on $\rho_{\star}$ is more constraining than the value of $\rho_{\star}$ one would obtain from the light curve, and because the inferred planetary parameters critically hinge upon $\rho_{\star}$, this ultimately leads to smaller uncertainties in the planetary parameters than we obtain purely from the light curve observables.

To show why this is true, we begin by comparing their prior in $\rho_{\star}$ (the value they derive from their estimate of $M_{\star}$ and $R_{\star}$, which we will denote $\rho_{\star, \rm prior}$) to the uncertainty in $\rho_{\star}$ from observables (denoted $\rho_{\star, \rm obs}$).

Their prior on $\rho_{\star}$ can be trivially calculated from $\rho_{\star} = 3M_{\star}/4\pi R_{\star}^{3}$, and its uncertainty, through propagation of error, is therefore simply 
\begin{equation}
   \bigg(\frac{\sigma_{\rho_{\star, \rm prior}}}{\rho_{\star, \rm prior}}\bigg)^{2} \approx
     \bigg(\frac{\sigma_{M_{\star}}}{M_{\star}}\bigg)^{2} +
    9\bigg(\frac{\sigma_{R_{\star}}}{R_{\star}}\bigg)^{2}.
\end{equation}

Inserting the appropriate values from \citet{Dragomir:2019} yields\footnote{We note that the actual value reported in Section 3.1 of \citet{Dragomir:2019} is $\rho_{\star} =3.09 \pm 0.23$ \dens, but after careful analysis, we believe that this value is probably a typographical error, as it differs from the value we derive and from the posterior value in Table 1 of the paper.} $\rho_{\star, \rm prior} = 3.07\pm 0.49$ \dens. This represents a fractional uncertainty of $\sigma_{\rho_{\star, \rm prior}}/\rho_{\star, \rm prior}=0.16$.

Now, combining Equations~\ref{eqn:arstar8} and~\ref{eqn:arstarobs}, we can express $\rho_{\star, \rm obs}$ and its uncertainty in terms of transit observables as
\begin{equation}
    \rho_{\star, \rm obs} = \bigg(\frac{3P}{G\pi^{2}}\bigg)\delta^{3/4} (T\tau)^{-3/2}.
    \label{eqn:rho_obs}
\end{equation}

Therefore, 
\begin{equation}
\begin{split}
     \bigg(\frac{\sigma_{\rho_{\star, \rm obs}}}{\rho_{\star, \rm obs}}\bigg)^{2} \approx
     \bigg(\frac{\sigma_P}{P}\bigg)^{2} +
    \frac{9}{16}\bigg(\frac{\sigma_{\delta}}{\delta}\bigg)^{2} + \frac{9}{4}\bigg(\frac{\sigma_{T}}{T}\bigg)^{2}+\\
    \frac{9}{4}\bigg(\frac{\sigma_{\tau}}{\tau}\bigg)^{2}.
    \label{eqn:error_rho_obs}
\end{split}
\end{equation}

Inserting the fractional uncertainties on $P$, $R_p$, $T$, and $\tau$ from the discovery paper into Equation~\ref{eqn:error_rho_obs}, we find  $\sigma_{\rho_{\star, \rm obs}}/{\rho_{\star, \rm obs}}=0.37$. This is larger and
less constraining than the fractional uncertainty in the prior on $\rho_{\star, \rm obs}$ from \citet{Dragomir:2019} by a factor of 2.3. Thus, we expect the prior on $\rho_{\star, \rm obs}$ to dominate over the constraint from the light curve. 
However, despite being considerably less constraining than the prior, the empirical constraint on $\rho_{\star, \rm obs}$ can still influence the posterior value if the central value is significantly different than the prior value. Inserting the values of $P$, $\delta$, $T$, and $\tau$ in Equation~\ref{eqn:rho_obs}, we find a central value of $\rho_{\star, \rm obs} = 5.56 \pm 2.06$ \dens. This value is $(5.56-3.07)/2.06=1.2\sigma$ discrepant from the prior value. Thus, there is a weak tension between the empirical and prior values of $\rho_{\star, \rm obs}$ that should be explored.

If we include the eccentricity in the expression for $\rho_{\star}$, we find much closer agreement between $\rho_{\star, \rm obs}$ and $\rho_{\star, \rm prior}$. 

From \citet{winn:2010}, we can express the scaled semi-major axis as a function of eccentricity as 
\begin{equation}
    \frac{a}{R_{\star}}= \frac{P}{\pi}\frac{\delta^{1/4}}{\sqrt{T\tau}} \bigg(\frac{\sqrt{1-e^{2}}}{1+e\sin{\omega}}\bigg).
    \label{eqn:arstarobseccentricity}
\end{equation}

We can then combine this equation with Equation~\ref{eqn:arstar8} to find the ratio between the inferred $\rho_{\star}$ assuming a circular orbit ($\rho_{\star,{\rm obs, c}}$) and that for an eccentric orbit ($\rho_{\star,{\rm obs, e}}$):
\begin{equation}
    \rho_{\star,{\rm obs, e}}=\rho_{\star,{\rm obs, c}} \bigg(\frac{\sqrt{1-e^{2}}}{1+e\sin{\omega}}\bigg)^3.
\end{equation}

Inserting the values from the paper ($e=0.188$ and $\omega=98^\circ$) yields $\rho_{\star,\rm obs} = 5.56$~\dens$\times 0.568=3.16$~\dens, and assuming the same fractional uncertainty as $\rho_{\star, \rm obs,c}$ of $0.37$ (which we discuss below), we get a value of $\rho_{\star, \rm obs,e} = 3.16\pm 1.17$ \dens, which is $\sim 0.1 \sigma$ greater than the prior, and in much better agreement than our estimate without including eccentricity. The reason why the eccentricity significantly affects $\rho_{\star, \rm obs}$ in this case, despite the fact that it is relatively small ($e= 0.188$), is that for this system, the argument of periastron is $\omega \simeq 90^{\circ}$, which implies that the transit occurs near periastron, and thus the transit is shorter than if the planet were on a circular orbit by a factor of 
\begin{equation}
    \frac{T_{\rm e}}{T_{\rm c}} \simeq  \frac{\sqrt{1-e^2}}{1+e} = 0.827. 
\end{equation}

Thus $\tau$ is shorter by the same factor. Since $\rho_{\star} \propto (T \tau)^{-3/2}$, by assuming $e=0$, one overestimates the density by factor of 
\begin{equation}
 \bigg(\frac{\sqrt{1-e^2}}{1+e}\bigg)^{-3} = 0.565^{-1},
\end{equation}
approximately recovering the factor above.

The eccentricity also affects the uncertainty in $\rho_{\star}$ in the following way:

\begin{equation}
\begin{split}
\frac{\rho_{\star,{\rm obs, e}}}
{\rho_{\star,{\rm obs, c}}}
   \propto \bigg(\frac{\sqrt{1-e^{2}}}{1+e\sin{\omega}}\bigg)^3 \simeq (1-3/2e^{2})(1-3e\sin{\omega})\\
    \simeq 1 - 3e\sin{\omega},
\end{split}
\end{equation}
where we have assumed that $e \ll 1$. Propagating the uncertainty leads to a final value of $\rho_{\star, \rm obs,e} = 3.16 \pm 2.04$ \dens, which is only $\sim$0.04$\sigma$ greater than $\rho_{\star, \rm prior}$. Thus, the eccentricity plays a significant role in the parameter uncertainties for this system. The uncertainty in the prior constraint on $\rho_{\star}$ is a factor of $\sim$1.7 times smaller than that derived from the data alone. We therefore conclude that the prior adopted by \citep{Dragomir:2019} dominates over the empirical value of $\rho_{\star}$ from the data ($\rho_{\star, \rm obs}$). This also explains why their final value and uncertainty in $\rho_{\star}$ ($3.03^{+0.50}_{-0.47}$ \dens) is so close to their prior ($\rho_{\star, \rm prior} = 3.07\pm 0.49$ \dens). 

Assuming that the uncertainty on their priors for $M_{\star}$ and $R_{\star}$ indeed dominates the fractional uncertainty in the resulting planet parameters, we can reproduce their uncertainties in $M_{p}$, $g_{p}$, and $\rho_{p}$ using their prior to recover their reported fractional uncertainties as follows.

For the surface gravity $g_{p}$, we have
\begin{equation}
    g_{p} = \frac{GM_{p}}{R_{p}^{2}}, 
\end{equation}

and
\begin{equation}
    M_{p} = \bigg(\frac{P}{2\pi G}\bigg)^{1/3} M_{\star}^{2/3} (1-e^2)^{1/2}K_{\star}
    \label{eqn:mass_prior}
\end{equation}

while the planet radius can be expressed as
\begin{equation}
    R_{p} = \delta^{1/2}R_{\star}
\end{equation}

Therefore,
\begin{equation}
    g_{p} = \bigg(\frac{P}{2\pi G}\bigg)^{1/3} M_{\star}^{2/3} K_{\star} \delta^{-1} R_{\star}^{-2} (1-e^{2})^{1/2}.
    \label{eqn:gpriors}
\end{equation}

Instead of simplifying $g_{p}$ in terms of observables (as we have done in Equation~\ref{eqn:gpobservables}), we express it in terms of $M_{\star}$ and $R_{\star}$. Using propagation of error, the uncertainty is 
\begin{equation}
\begin{split}
     \bigg(\frac{\sigma_{g_{p}}}{g_{p}}\bigg)^{2} \approx
     \frac{1}{9}\bigg(\frac{\sigma_P}{P}\bigg)^{2} +
    \frac{4}{9}\bigg(\frac{\sigma_{M_{\star}}}{M_{\star}}\bigg)^{2} + \bigg(\frac{\sigma_{K_{\star}}}{K_{\star}}\bigg)^{2}+\\
    \bigg(\frac{\sigma_{\delta}}{\delta}\bigg)^{2} + 4\bigg(\frac{\sigma_{R_{\star}}}{R_{\star}}\bigg)^{2}.
    \label{eqn:gpriorerror}
\end{split}
\end{equation}
where we have assumed $(1-e^{2})^{1/2} \approx 1$ since the eccentricity is small.

Inserting the appropriate values from Table 1 in \citet{Dragomir:2019} in Equations~\ref{eqn:gpriors} and~\ref{eqn:gpriorerror}, we recover a fractional uncertainty in the surface gravity of $\sigma_{g_p}/g_p = 0.14$, which is 12.5\% different from the value reported in \citet{Dragomir:2019}, and thus agrees much better with their results than our initial estimate.

For the planet's mass, we start from Equation~\ref{eqn:mass_prior} and propagate its uncertainty as
\begin{equation}
     \bigg(\frac{\sigma_{M_{p}}}{M_{p}}\bigg)^{2} \approx
     \frac{1}{9}\bigg(\frac{\sigma_P}{P}\bigg)^{2} +
    \frac{4}{9}\bigg(\frac{\sigma_{M_{\star}}}{M_{\star}}\bigg)^{2} + \bigg(\frac{\sigma_{K_{\star}}}{K_{\star}}\bigg)^{2},
    \label{eqn:Mppriorerror}
\end{equation}

implying a fractional uncertainty in the mass of $\sigma_{M_p}/M_p = 0.093$, which is only $\sim$0.3\% discrepant from the uncertainty in the paper. 

Finally, we replicate the analysis for the planet's density, starting with 

\begin{equation}
    \rho_p = \frac{3M_p}{4\pi R_p^3}=\frac{3M_p}{4\pi R_{\star}^3}\delta^{-3/2}.
    \label{eqn:rhop2}
\end{equation}

Therefore,
\begin{equation}
    \rho_{p} = \frac{3}{4\pi}\bigg(\frac{P}{2\pi G}\bigg)^{1/3} M_{\star}^{2/3} K_{\star} \delta^{-3/2} R_{\star}^{-3} (1-e^{2})^{1/2}.
\end{equation}

And the uncertainty in $\rho_{p}$ is thus
\begin{equation}
\begin{split}
     \bigg(\frac{\sigma_{\rho_{p}}}{\rho_{p}}\bigg)^{2} \approx
     \frac{1}{9}\bigg(\frac{\sigma_P}{P}\bigg)^{2} +
    \frac{4}{9}\bigg(\frac{\sigma_{M_{\star}}}{M_{\star}}\bigg)^{2} + \bigg(\frac{\sigma_{K_{\star}}}{K_{\star}}\bigg)^{2}+\\
    \frac{9}{4}\bigg(\frac{\sigma_{\delta}}{\delta}\bigg)^{2} + 9\bigg(\frac{\sigma_{R_{\star}}}{R_{\star}}\bigg)^{2},
\end{split}
\label{eqn:rhopGuenther}
\end{equation}

which leads to $\sigma_{\rho_{p}}/\rho_{p} = 0.18$, while the paper reports $\sigma_{\rho_{p}}/\rho_{p} = 0.21$, which is a $\sim$14\% difference. 

In summary, we can roughly reproduce the uncertainties in \citet{Dragomir:2019} to better than 15\% if we assume that such uncertainties are dominated by the priors on the stellar mass and radius.  In Figure~\ref{fig:fig1}, we plot both our initial fractional uncertainties and the recovered uncertainties as pairs connected by golden arrows that point in the direction of the `recovered' uncertainties based on our forensic analysis. 

\subsection{K2-106b}
\label{sec:k2106}

K2-106b is the inner planet in a system of two transiting exoplanets discovered by \citet{Adams:2017} and later characterized by \citet{Guenther:2017}. It is on an ultra short, 0.57 day orbit around a G5V star. It has a mass and radius of $8.36^{+0.96}_{-0.94} M_{\oplus}$ and $1.52\pm0.16 R_{\oplus}$, leading to a high bulk density of $\rho_{p} = 13.1^{+5.4}_{-3.6}$ \dens. \citet{Guenther:2017} used data from the K2 mission combined with multiple radial velocity observations from the High Dispersion Spectrograph (HDS; \citealt{Noguchi:2002}), the Carnegie Planet Finder Spectrograph (PFS; \citealt{Crane:2006}), and the FIber-Fed Echelle Spectrograph (FIES; \citealp{Frandsen:1999, Telting:2014}) to confirm and analyze this system. They performed a multi-planet joint analysis of the data using the code {\tt pyaneti} \citep{Barragan:2017} and derived the host star's mass and radius using the {\tt PARSEC} model isochrones and the interface for Bayesian estimation of stellar parameters from \citet{daSilva:2006}.

As with HD 21749b, we found large discrepancies ($\sim$50\%) between our analytic estimates of the uncertainties on the planetary mass and density and the literature values for K2-106b. Unlike HD 21749b, however, the reason for this discrepancy is that the uncertainty in the density of the host star, and thus in the properties of the planet, is dominated by the data, including the light curve + radial velocity, rather than the prior. To see why this is true, we perform a similar analysis as in Section~\ref{sec:hd21749b}, and begin by comparing the uncertainty in the density from the observables $\rho_{\star, \rm obs}$ to the density from the prior  $\rho_{\star, \rm prior}$. 

First, we have that the uncertainty in $\rho_{\star}$ based purely on the prior fractional uncertainties on $M_{\star}$ and $R_\star$ is given by
\begin{equation}
   \bigg(\frac{\sigma_{\rho_{\star, \rm prior}}}{\rho_{\star, \rm prior}}\bigg)^{2} \approx
     \bigg(\frac{\sigma_{M_{\star}}}{M_{\star}}\bigg)^{2} +
    9\bigg(\frac{\sigma_{R_{\star}}}{R_{\star}}\bigg)^{2}.
\end{equation}
Inserting the values of $\sigma_{M_\star}/M_\star$ and $\sigma_{R_\star}/R_\star$ from the paper, we derive a fractional uncertainty on the density of the star from the prior of $\sigma_{\rho_{\star,prior}}/\rho_{\star,prior}=0.31$. On the other hand, using Equation~\ref{eqn:error_rho_obs}, the fractional uncertainty in the stellar density from pure observables $\rho_{\star, \rm obs}$ is $\sigma_{\rho_{\star, \rm obs}}/\rho_{\star, \rm obs}=0.15$, a factor of $\sim$2 times smaller than the fractional uncertainty on $\rho_\star$ estimated from the prior. 

We can compute the uncertainty in the planetary mass assuming the fractional uncertainty on $M_{\star}$ from the prior and the fractional uncertainty on the measured semi-amplitude $K_{\star}$ using  Equation~\ref{eqn:Mppriorerror}:
\begin{equation}
     \bigg(\frac{\sigma_{M_{p}}}{M_{p}}\bigg)^{2} \approx
    \frac{4}{9}\bigg(\frac{\sigma_{M_{\star}}}{M_{\star}}\bigg)^{2} + \bigg(\frac{\sigma_{K_{\star}}}{K_{\star}}\bigg)^{2},
\end{equation}
where we have assumed that $\sigma_{P}/P \ll 1$.  We find $\sigma_{M_p}/M_p = 0.11$, which is only 1\% different from the uncertainty reported in \citet{Guenther:2017}. 

Further, we infer that \citet{Guenther:2017} estimated the density of the planet by combining their estimate of the mass of the planet by adopting the prior value of $M_\star$, along with the observed values of $K_\star$ and $P$, with the radius of the planet derived by adopting the prior value of $R_\star$ and the observed value of transit depth (and thus $R_p/R_\star$).  Thus we infer that \citet{Guenther:2017} estimated the  uncertainty in the planet density via,  
\begin{equation}
   \bigg(\frac{\sigma_{\rho_{p}}}{\rho_{p}}\bigg)^{2} \approx
     \bigg(\frac{\sigma_{M_{p}}}{M_{p}}\bigg)^{2} +
    9\bigg(\frac{\sigma_{R_{\star}}}{R_{\star}}\bigg)^{2} + \frac{9}{4}\bigg(\frac{\sigma_{\delta}}{\delta}\bigg)^{2},
\end{equation}
again assuming that $\sigma_{P}/P \ll 1$.  Substituting the values quoted in \citet{Guenther:2017} into the expression above, we find $\sigma_{\rho_{p}}/\rho_{p} = 0.32$, whereas they quote a fractional uncertainty of $\sigma_{\rho_{p}}/\rho_{p} = 0.34$, a $\sim$6\% difference. On the other hand, if we analytically estimate the fractional uncertainty on the density of K2-106b using pure observables (Equation~\ref{eqn:rhopuncertainty}), but assume their reported value and uncertainty on $R_{\star}$, we find  $\sigma_{\rho_{p}}/\rho_{p} \sim 0.18$, i.e., a factor of $\sim$2 times smaller. 

In the case of the surface gravity of the planet, however, we find that our analytic estimates and that reported in the paper only differ by $\sim 8\%$. The reason is in part because the uncertainty in the stellar density from the light curve dominates the uncertainty in the planet properties. In this case, the light curve and radial velocity data tightly constrain the stellar density, which implies that $M_{\star} \appropto R_{\star}^{3}$. This constraint on $\rho_{\star}$ causes the prior estimates of the stellar mass, radius, and their uncertainties to cancel out in the expression for the planet density:
\begin{equation}
    g_{p} \propto P^{1/3}K_{\star}M_{\star}^{2/3}\delta^{-1}R_{\star}^{-2}.
\end{equation}
Assuming $M_{\star} \propto \rho_\star R_\star^{3}$, and $\rho_{\star}\sim~{\rm constant}$, we find
\begin{equation}
    g_{p} \propto P^{1/3}K_{\star}R_{\star}^{2}\delta^{-1}R_{\star}^{-2} = P^{1/3}K_{\star}\delta^{-1}.
    \label{eqn:gpapprox}
\end{equation}

The reason why Equation~\ref{eqn:gpapprox} and Equation~\ref{eqn:gpobservables} do not agree is because Equation \ref{eqn:gpapprox} does not include the full contributions of the uncertainties in the light curve observables $P$, $\delta$, $T$, and $\tau$.  

Figure~\ref{fig:fig1} shows the fractional uncertainties in $M_{p}$, $g_{p}$, and $\rho_{p}$ for K2-106b and brown arrows pointing from the original values we estimate to the `recovered' values. 

We reanalyzed K2-106 with EXOFASTv2 to derive stellar and planetary properties without using either the MIST stellar tracks, the Yonsei Yale stellar evolutionary models (YY; \citealp{Yi:2001}), or the \citealp{Torres:2010} relationships that are built into EXOFASTv2. We first constrained the stellar radius by fitting the star's SED to stellar atmosphere models to infer the extinction $A_V$ and bolometric flux, which when combined with its distance from \textit{Gaia} EDR3 \citep{gaia:2020} provides a (nearly empirical) constraint on $R_{\star}$. We find a fractional uncertainty in $R_{\star}$ of 2.4\%, while \citet{Guenther:2017} derive a fractional uncertainty of 10\% using the measured values of $T_{\rm eff}$, $\log{g_\star}$, and [Fe/H] from their HARPS and HARPS-N spectra, combined with constraints from the {\tt PARSEC} model isochrones \citep{daSilva:2006}.

We used our estimate of the stellar radius to recalculate the fractional uncertainties in $M_{p}$, $g_{p}$, $\rho_{p}$, and $R_{p}$ using our analytic expressions, and the constraints on the empirical parameters $P,~K_{\star},~T,~\tau$, and $\delta$ from \citet{Guenther:2017}.  Our derived fractional uncertainty in the planetary radius is 3.0\%, whereas  \citet{Guenther:2017} find 10\% \footnote{We note that when $\sigma_{R_p/R_{\star}} \ll \sigma_{R_{\star}}/R_{\star}$, the fractional uncertainty on the planetary radius is equal to fractional uncertainty on the radius of the star (Eqn.~\ref{eqn:rpobs}). While this is approximately the case given fractional uncertainty in $R_{\star}$ estimated by \citet{Guenther:2017}, for our estimate the uncertainty in $R_p/R_{\star}$ of $1.9\%$ contributes somewhat to the our estimated fractional uncertainty in $R_p$.}. Our derived fractional uncertainty on the density of the planet is a factor of 2.3 times {\it smaller} than reported by \citep{Guenther:2017}.  This is because the radius of the star enters into their estimate of $\rho_p$ as $R_{\star}^{-3}$ (Eqn.~\ref{eqn:rhop2}), whereas our estimate of $\rho_p$ only depends linearly on $R_{\star}$ (Eqn.~\ref{eqn:rhopobs}). We estimate an uncertainty in the planet mass of 15\%, a bit larger than that reported by \citet{Guenther:2017}, as it scales as the square of the radius of the star (Eqn.~\ref{eqn:mpobs}). Finally, the planetary surface gravity uncertainty that we estimate is 14\%, almost the same as that estimated by \citet{Guenther:2017}, as it does not depend directly on $\sigma_{R_{\star}}/R_{\star}$.

We conclude that a careful reanalysis of the K2-106 system using purely empirical constraints may well result in a significantly more precise constraint on the density of K2-106b, which is already a strong candidate for an exceptionally dense super-Earth.  

\section{Discussion\label{sec:discussion}}

\begin{figure}
\begin{center}
\includegraphics[width=\columnwidth]{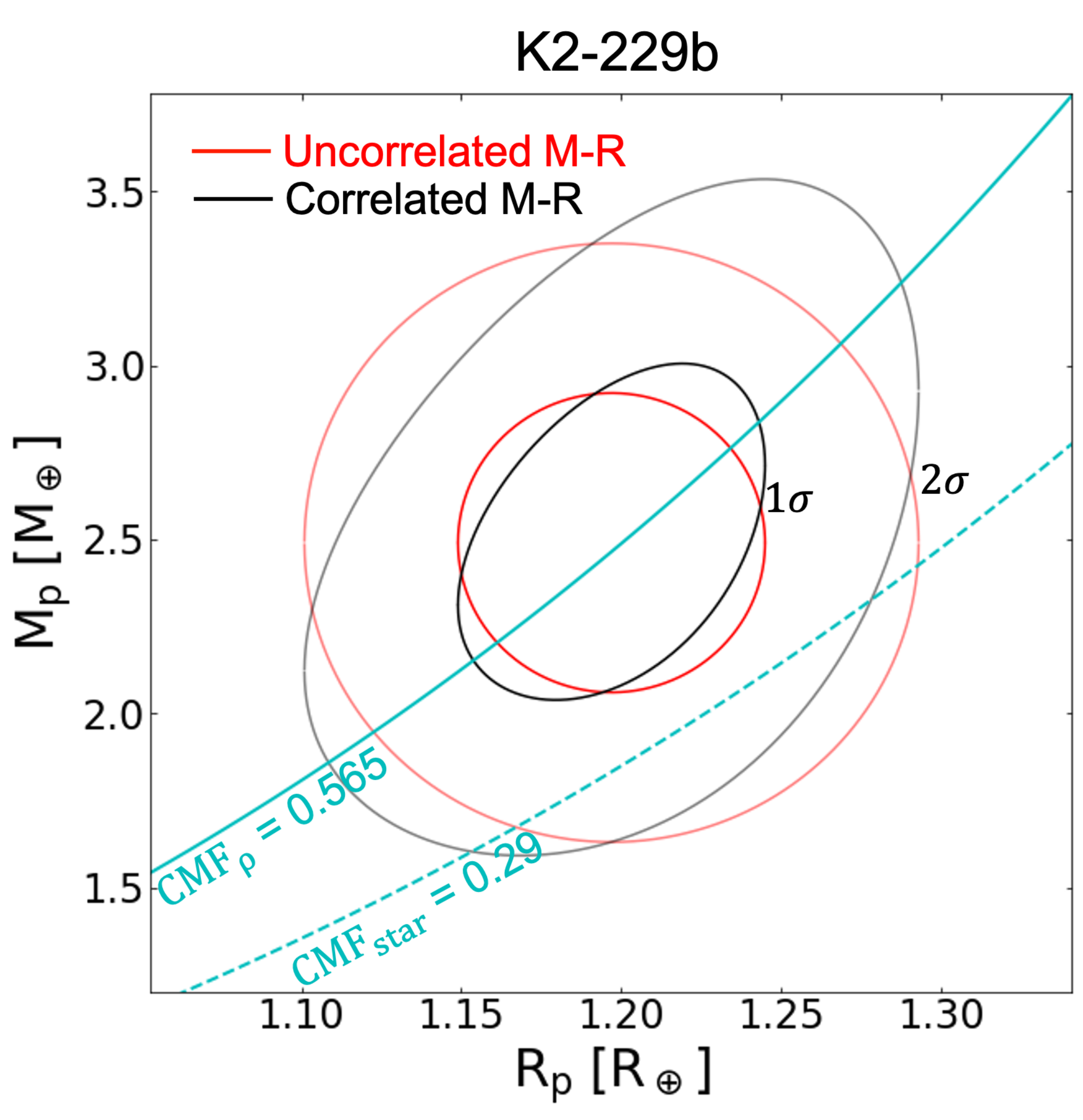}
\end{center}
\vspace{-3mm}
\caption{1$\sigma$ and 2$\sigma$ mass-radius ellipses for K2-229b \citep{Dai:2019}. The red ellipses assume $\rm M_{p}$ and $\rm R_{p}$ are uncorrelated, random variables. The black ellipses are the result of correlating $\rm M_{p}$ and $\rm R_{p}$ via the added constraint of surface gravity. Planets whose masses and radii lie along the blue solid solid line would have a constant core mass fraction of 0.565, whereas those that lie along the blue dotted line would have a constant core mass fraction of 0.29. Planets forming with iron abundances as expected from K2-229 Fe/Mg abundances will follow the blue dotted solid line.}
\label{fig:fig2}
\end{figure}

Here we discuss the importance of achieving high-precision measurements of planetary masses, surface gravities, densities and radii, and their overall role in a planet's habitability. 

The mass and radius of a planet are arguably its most fundamental quantities. The mass is a measure of how much matter a planet accreted during its formation and is also tightly connected to its density and surface gravity, which we discuss below. The mass also determines whether it can acquire and retain a substantial primordial atmosphere. Atmospheres are essential for a planet to maintain weather and thus life (see, e.g., \citealp{Dohm:2013}). In addition, the planetary core mass and radius (themselves a function of the total mass) are related to the strength of a planet's global magnetic field, although the strength of the field does depend on other factors, such as the rotation rate of the planet and other aspects of its interior. The presence of a substantial planetary magnetic field is vital in shielding against harmful electromagnetic radiation from the host star. This is especially true for exoplanets orbiting M dwarfs, which are much more active than Sun-like stars. Without a magnetic field to shield against magnetic phenomena such as flares and Coronal Mass Ejections (CMEs), planets around such stars may undergo mass loss and atmospheric erosion on relatively short timescales (see, e.g., \citealt{Kielkopf:2019}). The initial mass may also determine whether planets will have moons, a factor which has been hypothesized to play a role in the habitability of a planet, as it does for the Earth. Some authors have even proposed that Mars- and Earth-sized moons around giant planets may themselves be habitable (see, e.g., \citealp{Heller:2014,Hill:2018}).

The mean density of a planet is also important as it is a first-order approximation of its composition. Based on their density, we can classify planets as predominantly rocky (typically Earth-sized and super Earths) or gaseous (Neptune-sized and hot Jupiters). A reliable determination of the density and structure of a planet helps to constrain its habitability. 

Next, we briefly discuss a few aspects of the importance of knowledge of a planet's surface gravity. First, the surface gravity dictates the escape velocity of the planet, as well as the planet's atmospheric scale height, $h$, defined as

\begin{equation}
    h = \frac{k_{b}T_{\rm eq}}{\mu g_{p}},
\end{equation}

where $k_{b}$ is the Boltzmann constant, $T_{\rm eq}$ is the planet equilibrium temperature, and $\mu$ is the atmospheric mean molecular weight.  
The surface gravity is connected to mass loss events and the ability of a terrestrial planet to retain a secondary atmosphere. Perhaps most importantly, gravity may be a main driver of plate tectonics on a terrestrial planet. One of the most fundamental questions about terrestrial or Earth-like planets is whether they can have and sustain active plate tectonics or if they are in the stagnant lid regime, like Mars or Venus \citep{vanHeck:2011}. On Earth, plate tectonics are deeply linked to habitability for several crucial reasons. Plate tectonics regulate surface carbon abundance by transporting some $\rm CO_{2}$ out of the atmosphere and into the interior, which helps maintain a stable climate over long timescales \citep{Sleep:2001, unterborn:2016}. An excess of carbon dioxide can result in a runaway greenhouse effect, as in the case of Venus. Plate tectonics also drive the formation of surface features like mountains and volcanoes, and play an important role in sculpting the topography of a rocky planet. Weather can then bring nutrients from mountains to the oceans, contributing to the biodiversity of the oceans. Some authors have argued that plate tectonics, dry land (such as continents), and continents maximize the opportunities for intelligent life to evolve \citep{Dohm:2013}.

However, the origin and mechanisms of plate tectonics are poorly understood on Earth, and are even more so for exoplanets. The refereed literature on this topic includes inconsistent conclusions regarding the conditions required for plate tectonics, and in particular how the likelihood of plate tectonics depends on the mass of the planet. For example, there is an ongoing debate about whether plate tectonics are inevitable or unlikely on super Earths. \citet{valencia:2009} used a convection model and found that the probability and ability of a planet to harbor plate tectonics increases with planet size. On the other hand, \citet{oneill:2007} came to the opposite conclusion, finding that plate tectonics are less likely on larger planets, based on numerical simulations. The resolution to this debate will have important consequences for our assessment of the likelihood of life on other planets. 

\subsection{Surface gravity as a proxy for the core mass fraction}

The surface gravity of a planet may also play an important role in constraining other planetary parameters, like the core mass fraction. Here, we considered K2-229b ($R_{p} = 1.197^{+0.045}_{-0.048}~R_{\oplus}$ and $M_{p} =2.49 ^{+0.42}_{-0.43}~M_{\oplus}$, \citealt{Dai:2019}), a potential super Mercury first discovered by \citet{Santerne:2018}.  This planet has well measured properties and the prospects for improving the precision of the planet parameters are good given the brightness of the host star. 

We calculate the core mass fraction of K2-229b as expected from the planet's mass and radius, $\rm CMF_{\rho}$, which is the mass of the iron core divided by the total mass of the planet: $\rm CMF_{\rho}$ = $\rm M_{Fe}/M_{p}$. We compare this to the CMF as expected from the refractory elemental abundances of the host star, $\rm CMF_{star}$. This definition assumes that a rocky planet's mass is dominated by Fe and oxides of Si and Mg. Therefore, the stellar Fe/Mg and Si/Mg fractions are reflected in the planet's core mass fraction. The mass and radius of K2-229b are consistent with a rocky planet with a 0.57 core mass fraction (CMF), while the relative abundances of Mg, Si, and Fe of the host star K2-229 (as reported in \citealt{Santerne:2018}) predict a core mass fraction of 0.29 \citep{Schulze:2020}.  Figure~\ref{fig:fig2} shows mass-radius (M-R) ellipses for K2-229b when the mass and radius are assumed to be uncorrelated (red) and correlated via the added constraint of surface gravity (black). While apparently enriched in iron, the enrichment is only significant at the 2$\sigma$ level. The surface gravity, however, is correlated to the mass and gravity, reducing the uncertainty in $\rm CMF_{\rho}$ (black): the M-R ellipse that includes the surface gravity constraint reduces the uncertainty in the differences of CMF measures. This arises because the planet's density and surface gravity only differ by one factor of $R_{p}$. Because the black contours closely follow the line of constant 0.57 CMF, we assert that surface gravity and planet radius may be a better proxy for core mass fraction than mass and radius. Indeed, at the current uncertainties, we calculate that the additional constraint of surface gravity reduces the uncertainty in the $\rm CMF_{\rho}$ of K2-229b from 0.182 to 0.165. This is important given that we have demonstrated that the surface gravity of a planet is likely to be one of the most precisely measured properties of the planet. Furthermore, the fractional precision of the surface gravity measurement can be arbitrarily improved with additional data, at least to the point where systematic errors begin to dominate.

\section{Conclusions\label{sec:conclusions}}

One of the leading motivations of this paper was the answer to the question: ``given photometric and RV observations of a given exoplanet system, can we measure a planet's surface gravity better than its mass?" At first glance, the surface gravity depends on the mass itself, so it seems that the gravity should always be less constrained. However, upon expressing the mass, gravity and density as a function of photometric and RV parameters, we see that the mass and density have an extra dependence on the stellar radius, which makes the surface gravity generically easier to constrain to a given fractional precision than the mass or density. When expressed in terms of pure observables, a hierarchy in the precisions on the planet properties emerges, such that the surface gravity is better constrained than the density, and the latter is in turn better constrained than the mass. The surface gravity is a crucial planetary property, as it dictates the scale height of a planet's atmosphere. It is also a potential driver of plate tectonics, and as we show in this paper, can be an excellent proxy to constrain a planet's core mass fraction to better facilitate the discrimination of planet composition as different from its host star. With current missions like TESS, we expect to achieve high precisions in the photometric parameters. State-of-the-art RV measurements can now reach precisions in the semi-amplitude of $< 5\%$. As a result, the uncertainties in the ingress/egress duration $\tau$ and the host star radius $R_{\star}$ may be the limiting factors in constraining the properties of low-mass terrestrial planets.

\acknowledgements{We would like to thank Andrew Collier Cameron for his suggestion that the surface gravity of a transiting planet may be more well constrained than its mass, radius, or density.  R.R.M. and B.S.G. were supported by the Thomas Jefferson Chair for Space Exploration endowment from the Ohio State University. D.J.S. acknowledges funding support from the Eberly Research Fellowship from The Pennsylvania State University Eberly College of Science. The Center for Exoplanets and Habitable Worlds is supported by the Pennsylvania State University, the Eberly College of Science, and the Pennsylvania Space Grant Consortium. The results reported herein benefited from collaborations and/or information exchange within NASA’s Nexus for Exoplanet System Science (NExSS) research coordination network sponsored by NASA’s Science Mission Directorate. J.G.S. acknowledges the support of The Ohio State School of Earth Sciences through the Friends of Orton Hall research grant. W.R.P. was supported from the National Science Foundation under Grant No. EAR-1724693. 
}

\software{{\tt EXOFASTv2} \citep{Eastman:2017, eastman:2019}}.

\newpage
\bibliography{bibliography.bib}

\end{document}